\documentclass[aps,pre,floatfix]{revtex4}
\usepackage{epsf}
\usepackage{subfigure}
\usepackage{amsmath}
\usepackage{amssymb}
\usepackage{graphicx}
\usepackage{epstopdf}
\font\helvb=cmssbx12

\begin{document}

\newcommand{\be}{\begin{equation}}
\newcommand{\ee}{\end{equation}}
\newcommand{\bea}{\begin{eqnarray}}
\newcommand{\eea}{\end{eqnarray}}
\newcommand{\mean}[1]{\left \langle #1 \right \rangle}

\title{\bf Random paths and current fluctuations in nonequilibrium statistical mechanics}

\author{Pierre Gaspard}
\affiliation{Center for Nonlinear Phenomena and Complex Systems -- Department of Physics,\\
Universit\'e Libre de Bruxelles, Code Postal 231, Campus Plaine, 
B-1050 Brussels, Belgium}

\begin{abstract}
An overview is given of recent advances in nonequilibrium statistical mechanics about the statistics of random paths and current fluctuations.  Although statistics is carried out in space for equilibrium statistical mechanics, statistics is considered in time or spacetime for nonequilibrium systems.  In this approach, relationships have been established between nonequilibrium properties such as the transport coefficients, the thermodynamic entropy production, or the affinities, and quantities characterizing the microscopic Hamiltonian dynamics and the chaos or fluctuations it may generate.  This overview presents results for classical systems in the escape-rate formalism, stochastic processes, and open quantum systems.
\end{abstract}


\maketitle

\section{Introduction}
\label{Intro}

During the last decades, great advances have been carried out in nonequilibrium statistical mechanics with the discovery of dynamical large-deviation relationships underlying macroscopic transport properties and nonequilibrium thermodynamics.  For long, large-deviation relationships have been known in the formalism of equilibrium statistical mechanics where the physical quantities are defined per unit volume.  In the seventies, the development of chaos theory has led to the analogy between spin chains at equilibrium and time series in chaotic dynamics.  If the former systems extend in space, the latter ones evolve in time.  Otherwise, statistics is performed similarly in both cases by sampling random sequences of local properties.  In this way, a formalism has been developed by Bowen, Ruelle, and Sinai to characterize the dynamical properties of chaotic systems in terms of invariant probability measures \cite{S72,B75,R78}.  In this formalism, the analogue of the thermodynamic entropy per unit volume becomes the Kolmogorov-Sinai entropy per unit time, which characterizes temporal disorder also referred to as dynamical randomness in chaotic time series~\cite{K59,S59}.

More recently, these methods have been extended to nonequilibrium statistical mechanics on the ground of the same analogy.  Indeed, if equilibrium properties are stationary and, thus, time independent, nonequilibrium statistical mechanics deals with time-dependent properties.  Therefore, here also, extending statistics in space to statistics in time or spacetime proves useful.  If the early chaos-transport relationships set up conceptual frameworks to implement the large-deviation methods \cite{GN90,GB95,DG95,GD95}, the issue of time-reversal symmetry and its breaking in nonequilibrium steady states was addressed later on with the so-called fluctuation theorems and relationships to the thermodynamic entropy production \cite{Sardinia,ECM93,ES94,GC95,G96,K98,LS99,M99,J97,C98,C99,J00,J11,ES02,HS07,D07,S12,JOPS12}.  Nowadays, fluctuation theorems have been proved for all the current flowing across a nonequilibrium system would this latter be quantum or stochastic \cite{AG04,AG06JSM,AG07JSM,AG07JSP,FB08,SU08,AGMT09,EHM09,TH07,TLH07,CHT11}.  Moreover, such theorems have been shown to have implications not only in linear response theory where the Green-Kubo formulas and the Onsager reciprocity relations are recovered \cite{G52,K57,O31,C45}, but also in nonlinear response theory where the generalizations of these classic results have been discovered~\cite{AG04,AG07JSM,SU08,AGMT09,BK77,BK79}.

The purpose of the present paper is to present an overview of these advances in the framework of Hamiltonian classical or quantum dynamics, and the theory of stochastic processes.  Section~\ref{Chaos} summarizes work on chaos-transport relationships in open systems with escape.  Section~\ref{Paths} presents results on the connection between the thermodynamic entropy production and the  statistics of random paths and their time reversals.  Section~\ref{Currents} is devoted to the statistics of current fluctuations in stochastic and quantum systems and to the implications of the current fluctuation theorems.  Section~\ref{Equil} shows that broken symmetries in equilibrium states can also be characterized in terms of similar relationships.  Conclusions and perspectives are drawn in Section~\ref{Conclusions}.

\section{Chaos-transport relationships}
\label{Chaos}

\subsection{From equilibrium statistical mechanics to open dynamical systems}

Gibbs' probability measures of equilibrium statistical mechanics are weighting every microstate or spin configuration $\pmb{\sigma}\in\{+1,-1\}^{\Lambda}$ with $\Lambda\subset{\mathbb Z}^d$ by a Boltzmann factor according to
\be
p_{\pmb{\sigma}} = \frac{1}{Z} \, \exp\left(-\frac{E_{\pmb{\sigma}}}{k_{\rm B}T}\right) ,
\label{Gibbs}
\ee
where $E_{\pmb{\sigma}}$ is the energy of the microstate, $T$ the temperature, $k_{\rm B}$ Boltzmann's constant, and $Z$ the partition function to normalize the probability measure as $\sum_{\pmb{\sigma}}p_{\pmb{\sigma}}=1$.  The free energy is defined by $F=-k_{\rm B}T \ln Z$ in terms of the partition function.  Introducing the average energy $\langle E\rangle = \sum_{\pmb{\sigma}}p_{\pmb{\sigma}}E_{\pmb{\sigma}}$ and the thermodynamic entropy $S=-k_{\rm B} \sum_{\pmb{\sigma}}p_{\pmb{\sigma}}\ln p_{\pmb{\sigma}}$, the free energy is given by
\be
F = \langle E\rangle - T \, S \, ,
\label{F}
\ee
which is a basic thermodynamic relation for the canonical ensemble.  In the grand-canonical ensemble, the pressure $P$ multiplied by the volume $V$ would be given by $PV = -\langle E-N\mu\rangle + TS$ where $N$ is the number of particles and $\mu$ the chemical potential.

In deterministic dynamical systems, the analogue of a spin configuration $\pmb{\sigma}=\sigma_1\sigma_2\cdots\sigma_n$ in a one-dimensional chain is a sequence $\pmb{\omega}=\omega_1\omega_2\cdots\omega_n$ of phase-space cells $\omega_j$ visited by a trajectory at successive times $j\Delta t$ with $j=1,2,...,n$.  If the system is dynamically unstable with sensitivity to initial conditions, the probability weight given to the phase-space domain defined by the sequence $\pmb{\omega}$ decreases exponentially with $n$ at a rate controlled by the sum of local positive Lyapunov exponents $\beta\sum_{\lambda^{(i)}>0} \lambda_{\pmb{\omega}}^{(i)}$, which is at the basis of the analogy with Gibbs measures developed by Bowen, Ruelle, and Sinai:
\be
p_{\pmb{\omega}} \sim \exp\left( -\beta\sum_{\lambda^{(i)}>0} \lambda_{\pmb{\omega}}^{(i)} \, n \Delta t\right) ,
\label{p_beta}
\ee
where $\beta$ is an abstract parameter \cite{S72,B75,R78,ER85}.  An entropy per unit time is introduced as
\be
h= \lim_{n\to\infty} -\frac{1}{n\Delta t} \sum_{\pmb{\omega}} p_{\pmb{\omega}} \ln p_{\pmb{\omega}} \, ,
\label{dyn-entr}
\ee
which is associated with the partition of the phase space into the cells $\{\omega\}$.  The supremum over all the possible partitions defines the Kolmogorov-Sinai (KS) entropy per unit time $h_{\rm KS} = {\rm Sup}_{\{\omega\}} h$ \cite{K59,S59}.  This entropy is here associated with the probability measure (\ref{p_beta}) and, thus, depends on the parameter $\beta$.  The KS entropy characterizes the temporal disorder or dynamical randomness in the time evolution of the system.  The system is chaotic if its KS entropy is positive and non chaotic if it is vanishing.  This quantity is interpreted as the rate of production of information if the time evolution of the system would be observed by an apparatus and recorded in a digital memory.  The resolution of the apparatus corresponds to the size of the cells used to partition the phase space.  The supremum taken over all the possible partitions gives an estimation of the maximum accumulation rate of information in the memory that would be required to reconstruct typical trajectories of the system from the recorded data.  

Further quantities are introduced to obtain a relationship analogous to Eq.~(\ref{F}).  On the one hand,  the local Lyapunov are averaged over the probability measure (\ref{p_beta}) to get the mean Lyapunov exponents, $\langle\lambda^{(i)}\rangle_\beta$ \cite{ER85}.  On the other hand, Ruelle's topological pressure $P(\beta)$ is introduced as the generating function of the statistical moments of the sum of Lyapunov exponents.  The analogue of Eq.~(\ref{F}) reads
\be
P(\beta) = -\beta \sum_{\lambda^{(i)}>0} \langle\lambda^{(i)}\rangle_\beta + h_{\rm KS}(\beta) \, .
\label{Ruelle}
\ee

Among the different invariant probability measures~(\ref{p_beta}), the measure that is invariant under the time evolution ruled by Liouville's equation of statistical mechanics is given by the parameter value $\beta=1$.  The reason is that the classical dynamics stretches phase-space volumes by the factors $\exp\left( \lambda^{(i)}_{\pmb{\omega}} n\Delta t\right)$ in the unstable directions corresponding to the positive Lyapunov exponents and contracts them accordingly in the stable directions.  Liouville's theorem is satisfied because the sum of all the positive and negative Lyapunov exponents is vanishing $\sum_i \lambda^{(i)}_{\pmb{\omega}}=0$ so that phase-space volumes are preserved.

In a closed system where the total probability remains in a bounded phase-space domain, the invariant probability measure (\ref{p_beta}) with $\beta=1$ should precisely decay with the time $n\Delta t$ at the rate given by the sum of its positive Lyapunov exponents so that the KS entropy is thus equal to this sum:
\be
\mbox{closed systems:} \qquad h_{\rm KS} = \sum_{\lambda^{(i)}>0} \langle\lambda^{(i)}\rangle \, ,
\ee
where the parameter is here dropped since $\beta=1$.  

However, there exist open systems where the probability escapes to infinity.  This is the case in models of unimolecular chemical reactions where the Hamiltonian motion leads to the separation of the fragments of a molecule that is initially excited \cite{GR89}.  Most of the trajectories run to and from infinity, but some of them remain trapped forever in a bounded phase-space domain and they typically form a fractal set of unstable orbits.  The motion is transient in the vicinity of this fractal set, from which escape occurs.  If trajectories are launched from initial conditions near this fractal set, the number $N_t$ of trajectories remaining in the vicinity of this set decays exponentially with time $t\to\infty$ in systems of hyperbolic character. The escape rate is thus defined as $\gamma=\lim_{\rm t\to\infty} -(1/t)\ln N_t$.  A conditionally invariant probability measure can be defined on the fractal set by renormalizing the probability with the fraction that has escaped and the relation (\ref{Ruelle}) is obeyed.  Now, the KS entropy is no longer in balance with the sum of positive Lyapunov exponents at $\beta=1$ and their difference gives the escape rate according to the escape-rate formula
\be
\mbox{open systems:} \qquad \gamma = \sum_{\lambda^{(i)}>0} \langle\lambda^{(i)}\rangle - h_{\rm KS}
\label{esc_rate}
\ee
in systems of hyperbolic character \cite{ER85,GR89,KG85,CM97,CMT00}. The escape rate is given in terms of Ruelle's topological pressure by $\gamma=-P(1)$.

We notice that these methods can be extended to quantum systems for the study of quantum chaotic scattering~\cite{GR89}.

\subsection{The escape-rate formula and transport properties}

The contact with nonequilibrium statistical mechanics is established by considering first-passage problems in many-particle systems with transport by diffusion, viscosity or heat conductivity \cite{GN90,GB95,DG95,GD95,G98,D99}.  Let us consider a Hamiltonian system with $N$ particles in a finite volume $V$.  The Hamiltonian time evolution takes place in the phase space of the positions ${\bf r}_i=(x_i,y_i,z_i)$ and momenta ${\bf p}_i=(p_{ix},p_{iy},p_{iz})$ of the $N$ particles: ${\cal M}=\{({\bf r}_1,{\bf p}_1,...,{\bf r}_N,{\bf p}_N): \, {\bf r}_i\in V , \, {\bf p}_i\in{\mathbb R}^d, \, i=1,2,...,N\}$.  In this phase space, a hypersurface $\Sigma$ of dimension $6N-1$ is introduced, which is the border of a bounded domain on constant energy hypersurfaces.  A first-passage problem is set up at the hypersurface $\Sigma$.  Trajectories escape without return as soon as they reach the hypersurface $\Sigma$.  There may exist a set of trajectories trapped inside the hypersurface $\Sigma$, on which the probability measure~(\ref{p_beta}) could be constructed and the escape-rate formula (\ref{esc_rate}) would apply.

This construction has been carried out for open systems with diffusion such as Lorentz gases \cite{GN90,GB95}.  The transport property of diffusion can be considered in systems with independent particles (for which $N=1$).  Escape occurs as soon as the particle reaches the border $\Sigma$ of a domain delimited in position space. In Lorentz gases with a regular lattice, the macroscopic diffusion equation applies as proved by Bunimovich and Sinai in hard-disk billiards with a finite horizon and by Knauf in square lattices of Yukawa potentials \cite{BS80,K87}.  In these two-dimensional Lorentz gases,  there is a single positive Lyapunov exponent $\lambda$.  If the domain delimited by the border $\Sigma$ is large enough, the escape rate can be estimated by solving the diffusion equation with absorbing boundary conditions on the border $\Sigma$ of the domain.  If the domain is delimited by two parallel lines separated by the distance $L$, the escape rate is estimated as $\gamma\simeq {\cal D}(\pi/L)^2$ where $\cal D$ is the diffusion coefficient.  Therefore, combining with the escape-rate formula (\ref{esc_rate}), the diffusion coefficient is related to the characteristic quantities of chaos according to
\be
{\cal D}  = \lim_{L\to\infty} \left(\frac{L}{\pi}\right)^2 \big( \langle\lambda\rangle - h_{\rm KS}\big)_L \, ,
\label{esc-rate-D}
\ee
where the positive Lyapunov exponent and the KS entropy are defined over the fractal set of trajectories trapped between the boundaries separated by the distance $L$, which is thereafter taken to the limit $L\to\infty$ \cite{GN90,GB95,G98}.

These considerations can be generalized to the other transport properties by using the so-called Helfand moments \cite{DG95,GD95,G98,D99}.  These quantities are the centroids in physical space of the conserved quantities associated with every transport property.  The Helfand moment for shear viscosity is defined as
\be
G^{(\eta)} = \frac{1}{\sqrt{Vk_{\rm B}T}} \sum_{i=1}^N x_i \, p_{iy} \, ,
\ee
where $V$ is the volume of the system \cite{H60}.  We notice that the Helfand moment for diffusion is just $G^{({\cal D})}=x$.  The Helfand moments are known to undergo diffusive motion, here along the $x$-axis of physical space \cite{H60}.  This suggests to set up a first-passage problem for the Helfand moment.  The escape would occur if the phase-space trajectory would lead the Helfand moment outside the interval $-(\chi/2)\leq G^{(\eta)}\leq +(\chi/2)$ \cite{DG95}.  In this way, chaos-transport relationships similar to Eq.~(\ref{esc-rate-D}) have been obtained to the different transport coefficients \cite{VG03a,VG03b}.

We notice that this theory is based on the statistics of random paths in order to construct invariant probability measures using Eq.~(\ref{p_beta}).  Similar considerations apply to stochastic processes supposed to be obtained by coarse-graining the underlying Hamiltonian dynamics.  Stochastic Lorentz gases have been studied with these methods and, more recently, stochastic models of glasses for which a thermodynamics of histories or spacetime thermodynamics has been developed \cite{AvBED96,LAvW05,MGC05,GJLPvDvW07}.

\subsection{Hydrodynamic modes of relaxation}

In spatially extended periodic Lorentz gases, diffusion induces the decay of any perturbation with respect to a uniform density of particles.  At the macroscopic level of description, this decay is ruled by the diffusion equation, which is linear.  Therefore, an arbitrary solution can be decomposed into particular solutions that are periodic in space, such as $\exp(i {\bf k}\cdot{\bf r})$.  This mode has the spatial periodicity $\ell=2\pi/\Vert{\bf k}\Vert$ in the direction of the wave vector $\bf k$.  The decay of these diffusive modes is known to proceed at the rate $\gamma_{\bf k}={\cal D}{\bf k}^2$ in the long-time limit \cite{Ba75,BY80}.  They constitute the simplest examples of the hydrodynamic modes that may exist in fluids or solids and their understanding is fundamental to nonequilibrium statistical mechanics and condensed matter physics \cite{Ba75,BY80}.

In chaotic Lorentz gases on periodic lattices, it turns out that the diffusive modes can be constructed with the methods of dynamical systems theory at the Liouvillian level of description.  The diffusive modes of wave vector $\bf k$ can be defined in terms of a cumulative function taken over a one-dimensional set of initial conditions in phase space.  If this set is a circle of angular coordinate $\theta$, the cumulative function can be defined as
\be
F_{\bf k}(\theta) = \lim_{t\to\infty} \frac{\int_0^{\theta} d\theta' \, \exp\{i{\bf k}\cdot[{\bf r}_t(\theta')-{\bf r}_0(\theta')]\}}{\int_0^{2\pi} d\theta' \, \exp\{i{\bf k}\cdot[{\bf r}_t(\theta')-{\bf r}_0(\theta')]\}} ,
\label{modes}
\ee
where ${\bf r}_t(\theta)$ is the position of the particle issued from the initial condition $\theta$ \cite{GCGD01}.  This function is complex and depicts a fractal curve in the complex plane $({\rm Re}\, F_{\bf k}, {\rm Im}\, F_{\bf k})$.  Its Hausdorff dimension $D_{\rm H}({\bf k})$ depends on the wave vector $\bf k$ and tends to the unit value as the wave vector vanishes, i.e., as the diffusive mode converges towards the equilibrium uniform distribution.  Now, the Hausdorff dimension can be shown to be related to the mean positive Lyapunov exponent $\langle\lambda\rangle$ and to the diffusion coefficient by
\be
{\cal D} = \langle\lambda\rangle \, \lim_{{\bf k}\to 0} \frac{D_{\rm H}({\bf k})-1}{{\bf k}^2} ,
\label{D-k}
\ee
as shown in Ref.~\cite{GCGD01}.  Although the framework where the diffusive modes~(\ref{modes}) are constructed is different from the escape-rate formalism, Eq.~(\ref{D-k}) is reminiscent of the escape-rate formula (\ref{esc-rate-D}) because fractal dimensions appear as proportionality factors between positive Lyapunov exponents and the KS entropy in open systems.  The fractal structure of the diffusive modes controls the long-time relaxation towards equilibrium and the thermodynamic entropy production in this class of systems \cite{DGG02}.

These chaos-transport relationships are based on the large-deviation properties of the microscopic dynamics, i.e., on the statistics of random paths generated by the Hamiltonian motion in phase space.

\section{Statistics of random paths}
\label{Paths}

\subsection{Stochastic processes}

Now, we consider systems in nonequilibrium steady states. In order to construct such states, the system should be in contact with large reservoirs and exchange particles or energy with them.  The reservoirs are themselves physical systems composed of particles moving according to Hamiltonian dynamics so that the reservoirs have necessarily more degrees of freedom than the system itself.  All these degrees of freedom not only induce thermodynamic forces driving the system out of equilibrium, but also concomitant noises in the form of thermal or molecular fluctuations.

The total system including the reservoirs can be treated starting from the Hamiltonian description of all its degrees of freedom.  In principle, such a treatment can be carried out in classical or quantum systems.  Often, the system of interest admits a mesoscopic description in terms of a stochastic process.  For instance, a Brownian particle driven out of equilibrium in a moving optical trap is well described by a stochastic Langevin equation, which can be validated by direct experimental measurements \cite{W54}.  Such stochastic descriptions have been experimentally validated for other processes down to the nanoscale \cite{G10}.  Moreover, stochastic processes can also be set up for the study of transport properties in systems of Hamiltonian type, in particular, heat conductivity and Fourier's law in many-particle billiards~\cite{GG08}.

If well-defined discrete states $\{\omega\}$ can be identified in the system under observation, the time evolution could be described as a time-continuous jump process ruled by the master equation
\be
\frac{dp(\omega,t)}{dt} = \sum_{\omega'(\neq\omega)} \left[ p(\omega',t) \, W(\omega',\omega) - p(\omega,t) \, W(\omega,\omega')\right]
\label{master}
\ee
for the probability $p(\omega,t)$ that the system is observed in the coarse-grained state $\omega$ at the time $t\in{\mathbb R}$ \cite{N72,NP77,S76,vK81,G04,W95}.  The quantity $W(\omega',\omega)$ is the rate of the transition $\omega'\to\omega$, i.e., the number of these transitions per unit time.  In this framework, the thermodynamic entropy has been identified for long to be given by
\be
S= \sum_\omega S^0(\omega) \, p(\omega,t) - k_{\rm B} \sum_\omega p(\omega,t) \, \ln p(\omega,t) \, ,
\label{S}
\ee
where $S^0(\omega)$ is the entropy if the system stays in the coarse-grained state $\omega$ and the second term is the contribution of the probability distribution $p(\omega,t)$ over the different possible coarse-grained states observed at the time $t$ \cite{G04JPC}.  The time derivative of Eq.~(\ref{S}) can be carried out with the master equation (\ref{master}) and is known to split in two contributions as
\be
\frac{dS}{dt} = \frac{d_{\rm e}S}{dt} +\frac{d_{\rm i}S}{dt} ,
\ee
where $d_{\rm e}S/dt$ is the entropy flow, which can be positive or negative, and 
\be
\frac{d_{\rm i}S}{dt} = \frac{k_{\rm B}}{2} \sum_{\omega\neq\omega'} \left[ p(\omega,t) \, W(\omega,\omega') - p(\omega',t) \, W(\omega',\omega)\right] \, \ln \frac{p(\omega,t) \, W(\omega,\omega')}{p(\omega',t) \, W(\omega',\omega)} \geq 0
\label{entr-prod}
\ee
is the entropy production, which is always non negative in agreement with the second law of thermodynamics~\cite{G04JPC}.  In the macroscopic limit, this expression is known to give the standard expressions of entropy production in hydrodynamics and chemical kinetics \cite{N72,NP77,S76,vK81,G04,W95}.  In nonequilibrium steady states, the entropy production (\ref{entr-prod}) is positive, although it vanishes at equilibrium where the detailed balance conditions hold, according to which $p_{\rm eq}(\omega) \, W(\omega,\omega') = p_{\rm eq}(\omega') \, W(\omega',\omega)$ for every transition $\omega\leftrightharpoons \omega'$.

\subsection{Random paths of stochastic processes}

The stochastic process (\ref{master}) generates random paths, in which the system is found in the state $\omega_k$ during the time interval $t\in[t_{k-1},t_k]$.  The dwell times $t_k-t_{k-1}$ are exponentially distributed with the average dwell time given by $\langle t_k-t_{k-1}\rangle^{-1} = \sum_{\omega'(\neq\omega_k)}W(\omega_k,\omega')$.  The jumps between the discrete states happen at the times $\{t_k\}_{k\in{\mathbb Z}}$.  If the process is observed with a stroboscope at regular time intervals $\Delta t$, random sequences $\pmb{\omega}=\omega_1\omega_2\cdots\omega_n$ are recorded where $\omega_j$ is the discrete state observed at the time $t=j\Delta t$.  The probability to observe such a sequence is given by
\be
p_{\pmb{\omega}} = p_{\omega_1} \, P(\omega_1,\omega_2) \, \cdots \, P(\omega_{n-1},\omega_n) \, ,
\label{path-prob}
\ee
where $p(\omega_j)$ is the stationary probability of the state $\omega_j$ and
$P(\omega_{j-1},\omega_j)$ is the conditional probability that the system is found in the state $\omega_j$ at the time $t=j\Delta t$ provided that it was in the state $\omega_{j-1}$ at the time $t=(j-1)\Delta t$.  This conditional probability is given in terms of the matrix $\mbox{\helvb W}$ of the transition rates according to
\be
P(\omega,\omega') = \left[ \exp\left( \mbox{\helvb W} \, \Delta t\right)\right]_{\omega\omega'} \, .
\ee
The dynamical randomness of the paths can be characterized by the entropy per unit time (\ref{dyn-entr}), which here depends on the sampling time $\Delta t$ because the process is continuous in time.  In the limit where $\Delta t\to 0$, the entropy per unit time behaves as \cite{G98,GW93,G04JSP}
\be
h(\Delta t) = \left(\ln\frac{\rm e}{\Delta t}\right) \sum_{\omega\neq\omega'} p(\omega) \, W(\omega,\omega') - \sum_{\omega\neq\omega'} p(\omega) \, W(\omega,\omega') \, \ln W(\omega,\omega') + O(\Delta t) \, .
\label{h-dt}
\ee
This quantity represents the accumulation rate of information needed to record a typical random path of the process with the sampling $\Delta t$.  The smaller the sampling time, the larger the rate because randomness is generated continuously in time.  Such stochastic processes are much more random than chaotic dynamical systems since they are characterized by an infinite KS entropy reached in the limit $\Delta t\to 0$.

\subsection{Broken time-reversal symmetry in nonequilibrium steady states}

The time-reversal symmetry of the process can be investigated by defining the average decay rate of the probability to observe the time reversal of some path
\be
\pmb{\omega}^{\rm R} = \omega_n \cdots \omega_2\omega_1
\ee
among the typical paths of the process \cite{G04JSP,MN03}.  This average decay rate is defined by an expression similar to Eq.~(\ref{dyn-entr}) 
\be
h^{\rm R}= \lim_{n\to\infty} -\frac{1}{n\Delta t} \sum_{\pmb{\omega}} p_{\pmb{\omega}} \ln p_{\pmb{\omega}^{\rm R}} \, .
\label{dyn-coentr}
\ee
The difference between both quantities is given by
\be
h^{\rm R}-h = \lim_{n\to\infty} \frac{1}{n\Delta t} \, D(p_{\pmb{\omega}} \Vert p_{\pmb{\omega}^{\rm R}}) \label{coentr-entr}
\ee
in terms of a relative entropy or Kullback-Leibler divergence \cite{W78,CT06}, which is always non negative
\be
D(p_{\pmb{\omega}} \Vert p_{\pmb{\omega}^{\rm R}}) = \sum_{\pmb{\omega}} p_{\pmb{\omega}}  \ln \frac{p_{\pmb{\omega}}}{p_{\pmb{\omega}^{\rm R}}} \geq 0 \, .
\label{KL-div}
\ee
In this regard, the quantity (\ref{dyn-coentr}) could be called a {\it coentropy} since it combines with the entropy to form the non-negative Kullback-Leibler divergence \cite{G12JSM}.  If the process is time-reversal symmetric and the probabilities of every path and its time reversal are equal, $p_{\pmb{\omega}} =p_{\pmb{\omega}^{\rm R}}$, the coentropy is equal to the entropy per unit time and the Kullback-Leibler divergence vanishes.  Therefore, this divergence characterizes the time asymmetry of the stochastic process.

If the coentropy is calculated with the path probability (\ref{path-prob}), we get
\be
h^{\rm R}(\Delta t) = \left(\ln\frac{\rm e}{\Delta t}\right) \sum_{\omega\neq\omega'} p(\omega) \, W(\omega,\omega') - \sum_{\omega\neq\omega'} p(\omega) \, W(\omega,\omega') \, \ln W(\omega',\omega) + O(\Delta t) \, ,
\label{hR-dt}
\ee
which is similar to Eq.~(\ref{h-dt}) except a permutation of $\omega$ and $\omega'$ in the transition rate appearing in the logarithm \cite{G04JSP}. Remarkably, the difference between the coentropy and the entropy converges in the limit $\Delta t\to 0$ to the thermodynamic entropy production (\ref{entr-prod}) of the nonequilibrium steady state:
\be
\frac{1}{k_{\rm B}} \, \frac{d_{\rm i}S}{dt} = \lim_{\Delta t\to 0} \left[ h^{\rm R}(\Delta t)- h(\Delta t)\right] \geq 0 \, .
\label{entr=hR-h}
\ee
This relation shows that dynamical order manifests itself in nonequilibrium systems \cite{G07CRP}.  Related results have been obtained for systems driven by time-dependent external forces \cite{KPV07}.

A few comments are here in order about the similarities with respect to chaos-transport relationships such as Eq.~(\ref{esc-rate-D}).  Both types of relationships share the common structure that they express an irreversible property as the difference between two large-deviation dynamical quantities characterizing the random paths of the process on a more microscopic scale.  Furthermore, we notice the similitude of both quantities appearing in this difference where the first one is an average while the second one is an non-negative entropy per unit time.  Since this latter characterizes dynamical randomness in both contexts, these relationships show that there is in general no proportionality between dynamical randomness and irreversible properties.  The rate of information production by the microscopic dynamics or a corresponding stochastic process is typically very large compared to the rate of thermodynamic entropy production, which may even vanish at equilibrium although random transitions continue to proceed because of thermal agitation and molecular fluctuations.  In this regard, the difference is needed to establish the connection.  

However, the relations (\ref{esc-rate-D}) and (\ref{entr=hR-h}) apply to distinct situations. The escape-rate formula (\ref{esc-rate-D}) concerns the internal classical dynamics of the system on the trapped trajectories, which form a set of zero Lebesgue measure in phase space.  In contrast, the relation (\ref{entr=hR-h}) concerns the complementary set of trajectories coming from one reservoir and going to the other one because this set constitutes the support of the nonequilibrium steady state.  Moreover, the latter relation characterizes the breaking of the time-reversal symmetry since the coentropy (\ref{dyn-coentr}) combines with the dynamical entropy (\ref{dyn-entr}) to form the Kullback-Leibler divergence (\ref{KL-div}) that vanishes when the symmetry is restored at equilibrium.  The relation (\ref{entr=hR-h}) has been tested experimentally in nonequilibrium Brownian motion and electric $RC$ circuits \cite{AGCGJP07,AGCGJP08}.

\subsection{Effusion}

\begin{figure}[h]
\begin{center}
\includegraphics[scale=0.35]{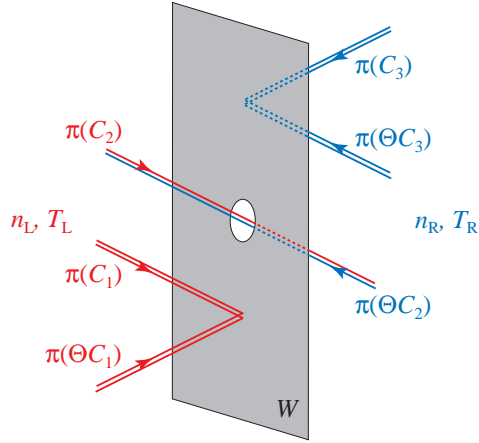}
\caption{Schematic representation of the effusion process in the position space ${\bf r}=(x,y,z)\in({\mathbb R}^3\setminus W)$ where $W$ denotes the wall between the two reservoirs.  The gas of non-interacting particles has different temperatures and densities in these reservoirs.  The particles may flow between the reservoirs through a hole of area $\sigma$ in the wall $W$.  The figure depicts the projection $\pi({\bf r},{\bf p})={\bf r}$ of three types of phase-space orbits $C_i$ and their time reversal $\Theta C_i$ (with $i=1,2,3$).}
\label{fig1}
\end{center}
\end{figure}

As an interesting example, we may consider the effusion process of a dilute gas through a small hole of area $\sigma$ in a wall separating two reservoirs at different temperatures and particle densities, $(n_{\rm L}, T_{\rm L})$ and $(n_{\rm R}, T_{\rm R})$ \cite{K1909,P58,CVK06,GA11}.  This process is illustrated in Fig.~\ref{fig1}.  The ideal gas is composed of monoatomic particles of mass $m$ moving in free flights, which are the solutions of Hamilton's equations for the single-particle position ${\bf r}\in{\mathbb R}^3$ and momentum ${\bf p}\in{\mathbb R}^3$ with the Hamiltonian function given by $H = {\bf p}^2/(2m)$. The free flights are eventually interrupted by elastic collisions on the thin wall $W$ reflecting the $z$-component of the momentum according to the collision rule, ${\bf p}'={\bf p}-2\,({\bf p}\cdot{\bf u}_z)\,{\bf u}_z$, where ${\bf u}_z$ is the unit vector in the $z$-direction perpendicular to the thin wall $W$.  The motion of every particle is thus ruled by a Hamiltonian flow $\Phi^t$ in the single-particle phase space ${\cal M}=\{({\bf r},{\bf p})\in({\mathbb R}^3\setminus W)\otimes{\mathbb R}^3\}$.  This single-particle flow is thus symplectic, obeys Liouville's theorem, and is symmetric under the time-reversal transformation $\Theta({\bf r},{\bf p})=({\bf r},-{\bf p})$: $\Theta\circ\Phi^t\circ\Theta=\Phi^{-t}$.

In the left-hand reservoir, the particles come from $z=-\infty$ with velocities distributed according to a Maxwell-Boltzmann function at the temperature $T_{\rm L}$ and the density $n_{\rm L}$.  In the right-hand reservoir, they come from $z=+\infty$ with velocities distributed at the temperature $T_{\rm R}$ and the density $n_{\rm R}$.  The single-particle distribution function is thus given by
\be
F({\bf r},{\bf p}) = \frac{n_C}{(2\pi mk_{\rm B} T_C)^{3/2}} \, \exp\left(-\frac{{\bf p}^2}{2mk_{\rm B} T_C}\right) \, ,
\label{distrib_fn}
\ee
where the values of the particle density $n_C$ and the temperature $T_C$ are associated with the orbit $C$ to which the phase-space point $({\bf r},{\bf p})$ belongs and corresponding to the domain from which the orbit is coming.  For the three types of possible orbits depicted in Fig.~\ref{fig1}, these values are the following:
\bea
&& n_C=n_{\rm L}\, , \quad T_C=T_{\rm L} \qquad\,\mbox{if}\qquad ({\bf r},{\bf p}) \in C_1\, , \ \Theta C_1 \, , \ C_2 \, ; \label{C=L}\\
&& n_C=n_{\rm R}\, , \quad T_C=T_{\rm R} \qquad\mbox{if}\qquad ({\bf r},{\bf p}) \in C_3\, , \ \Theta C_3 \, , \ \Theta C_2 \, . \label{C=R}
\eea
The distribution function is invariant under the Hamiltonian flow, $F[\Phi^t({\bf r},{\bf p})]=F({\bf r},{\bf p})$.
However, it is not symmetric under time reversal, $F[\Theta({\bf r},{\bf p})]\neq F({\bf r},{\bf p})$, for nonequilibrium constraints such that $n_{\rm L}\neq n_{\rm R}$ or $T_{\rm L}\neq T_{\rm R}$.

The effusion of the gas from each reservoir through the hole can be described as a stochastic process ruled by the master equation:
\bea
\frac{d}{dt}\, p_t(\Delta E,\Delta N) &=& \int_0^{\infty} d\epsilon\, 
w_{\rm L}(\epsilon) \left[ p_t(\Delta E-\epsilon,\Delta N-1)- p_t(\Delta E,\Delta N)\right] \nonumber\\
&& + \int_0^{\infty} d\epsilon\, w_{\rm R}(\epsilon) \left[ p_t(\Delta E+\epsilon,\Delta N+1) - p_t(\Delta E,\Delta N)\right] ,
\label{master-eff}
\eea
for 
the probability $p_t(\Delta E,\Delta N)$ that an energy $\Delta E$ and $\Delta N$ particles are transferred in the direction ${\rm L}\to{\rm R}$ during the time interval $t$ \cite{CVK06}.  The transition rates are given by
\bea
w_{\rm L}(\epsilon) &=& 
\frac{\sigma\, n_{\rm L}}{\sqrt{2\pi m k_{\rm B}T_{\rm L}}} \, \frac{\epsilon}{k_{\rm B}T_{\rm L}} \, \exp\left(-\frac{\epsilon}{k_{\rm B}T_{\rm L}}\right) \, , \label{rate-L}\\
w_{\rm R}(\epsilon) &=& 
\frac{\sigma\, n_{\rm R}}{\sqrt{2\pi m k_{\rm B}T_{\rm R}}} \, \frac{\epsilon}{k_{\rm B}T_{\rm R}} \, \exp\left(-\frac{\epsilon}{k_{\rm B}T_{\rm R}}\right)\, , \label{rate-R}
\eea
for particles of kinetic energy $\epsilon={\bf p}^2/(2m)$. This stochastic process is non stationary. 

The system is driven out of equilibrium by the thermodynamic force, also called affinities defined as
\bea
\mbox{thermal affinity:}\qquad&& A_E \equiv \frac{1}{k_{\rm B}T_{\rm R}}-\frac{1}{k_{\rm B}T_{\rm L}} \, , \label{AE}\\
\mbox{chemical affinity:}\qquad&& A_N \equiv \frac{\mu_{\rm L}}{k_{\rm B}T_{\rm L}}-\frac{\mu_{\rm R}}{k_{\rm B}T_{\rm R}} \, , \label{AN}
\eea
where $\mu_{\rm L}$ and $\mu_{\rm R}$ are the chemical potentials of the reservoirs \cite{DD36,P67,C85}.  Since the gas is supposed to be ideal and monoatomic, the affinity of the particle flow is given by
\be
A_N = \ln\left[ \frac{n_{\rm L}}{n_{\rm R}}\left(\frac{T_{\rm R}}{T_{\rm L}}\right)^{3/2}\right] \, .
\ee
We notice that the transition rates (\ref{rate-L})-(\ref{rate-R}) are related to each other and to the affinities (\ref{AE})-(\ref{AN}) by
\be
\frac{w_{\rm L}(\epsilon)}{w_{\rm R}(\epsilon)} = \exp\left( \epsilon A_E + A_N \right) ,
\label{W-ratio}
\ee
as the consequence of the nonequilibrium constraints on the flow of particles between both reservoirs.
At the equilibrium thermodynamic state where the affinities vanish, the two transition rates are equal
and the conditions of detailed balancing are recovered.   In the long-time limit, the thermodynamic entropy production is known to be given by \cite{CVK06}
\be
\frac{1}{k_{\rm B}} \, \frac{d_{\rm i}S}{dt} = A_E \langle J_E\rangle + A_N \langle J_N\rangle \geq 0
\ee
in terms of the affinities (\ref{AE})-(\ref{AN}) and the average values of the net energy and particle currents between the reservoirs:
\bea
&& \langle J_E\rangle = \int_0^{\infty} d\epsilon \, \epsilon \left[ w_{\rm L}(\epsilon)-w_{\rm R}(\epsilon)\right] , \label{JE}\\
&& \langle J_N\rangle  = \int_0^{\infty} d\epsilon \left[ w_{\rm L}(\epsilon)-w_{\rm R}(\epsilon)\right] . \label{JN}
\eea

Because of the infinite spatial extension of the reservoirs, the ideal gas contains an infinite number of particles and an invariant probability measure can be constructed as a Poisson suspension on the basis of the invariant distribution function (\ref{distrib_fn}) \cite{CFS82}.  The single-particle distribution function gives the density of particles in an element of volume $d{\bf r}\,d{\bf p}$ at some point $({\bf r},{\bf p})$ in the single-particle phase space ${\cal M}$.  Accordingly, a microstate of the infinite-particle system is given by
\be
\pmb{\Gamma} = ({\bf r}_1,{\bf p}_1,{\bf r}_2,{\bf p}_2,...,{\bf r}_N,{\bf p}_N,...)\in\pmb{\mathcal M}={\mathcal M}^{\infty} \, .
\label{microstate}
\ee
The Hamiltonian flow $\pmb{\Gamma}_t=\pmb{\Phi}^t(\pmb{\Gamma} _0)$ of this infinite system is also time-reversal symmetric:
\be
\pmb{\Theta}\circ\pmb{\Phi}^t\circ\pmb{\Theta}=\pmb{\Phi}^{-t} \, ,
\label{microreversibility}
\ee
where $\pmb{\Theta}$ is the time-reversal transformation acting on the microstate (\ref{microstate}) by reversing all the momenta: ${\bf p}_i\to -{\bf p}_i$, $\forall \, i \in {\mathbb N}$.

The Poisson suspension is constructed as follows.  For any six-dimensional domain ${\mathcal D}\subset{\mathcal M}$ of the single-particle phase space $\mathcal M$, we consider the random events 
\be
{\mathcal A}_{{\mathcal D},N} = \left\{ \pmb{\Gamma}\in\pmb{\mathcal M}: \; {\rm card}(\pmb{\Gamma}\cap{\cal D})=N\right\} \, ,
\ee
for which there are $N$ particles in the domain $\mathcal D$.  The average number of particles in the phase-space domain $\mathcal D$ is given by
\be
\nu({\mathcal D}) = \int_{\mathcal D} F({\bf r},{\bf p}) \, d{\bf r} \, d{\bf p}
\label{nu}
\ee
in terms of the single-particle distribution function (\ref{distrib_fn}).  
The number $N$ of particles in the domain $\mathcal D$ is a random variable of Poisson distribution:
\be
P\left({\mathcal A}_{{\mathcal D},N}\right) = \frac{\nu({\mathcal D})^N}{N!} \, {\rm e}^{-\nu({\mathcal D})} \, .
\label{Poisson}
\ee
Moreover, random events in which disjoint domains have given particle numbers, are statistically independent so that
\be
P\left({\mathcal A}_{{\mathcal D}_1,N_1}\cap{\mathcal A}_{{\mathcal D}_2,N_2}\right) = P\left({\mathcal A}_{{\mathcal D}_1,N_1}\right) \; P\left({\mathcal A}_{{\mathcal D}_2,N_2}\right) \qquad\mbox{if}\qquad {\mathcal D}_1\cap{\mathcal D}_2=\emptyset \, .
\label{indepstat}
\ee
Both Eqs.~(\ref{Poisson}) and (\ref{indepstat}) define the probability distribution of the so-called Poisson suspension \cite{CFS82}.  The measure $P$ is normalized to unity and, therefore, defines a probability measure.  This probability measure is invariant under the time evolution of the Hamiltonian flow $\pmb{\Phi}^t$:
\be
P\left(\pmb{\Phi}^t{\mathcal A}\right) = P\left({\mathcal A}\right)
\ee
for any random event $\mathcal A$.  This is the consequence of the stationarity of the single-particle distribution function (\ref{distrib_fn}), $F[\Phi^t({\bf r},{\bf p})]=F({\bf r},{\bf p})$, which implies the invariance $\nu\left(\Phi^t{\mathcal D}\right)=\nu\left({\mathcal D}\right)$ of the measure (\ref{nu}).  The flow $\pmb{\Phi}^t$ in the infinite phase space $\pmb{\mathcal M}$ and the invariant probability measure of the Poisson suspension defines the infinite-particle dynamical system $(\pmb{\Phi}^t,\pmb{\mathcal M},P)$. This dynamical system is known to have the ergodic, mixing, and Bernoulli properties \cite{CFS82}. Although the flow $\pmb{\Phi}^t$ has the microreversibility (\ref{microreversibility}), the dynamical system  is not symmetric under time reversal in general because
\be
P\left(\pmb{\Theta}{\mathcal A}\right) \neq P\left({\mathcal A}\right) \qquad\mbox{if}\qquad n_{\rm L} \neq n_{\rm R} \quad\mbox{or}\quad T_{\rm L} \neq T_{\rm R} \, ,
\ee
as noticed in Ref.~\cite{G98}.  The time-reversal symmetry is thus broken at the statistical level of description for the stationary probability distribution $P$ under nonequilibrium conditions $n_{\rm L}\neq n_{\rm R}$ or $T_{\rm L}\neq T_{\rm R}$.

The dynamical randomness can be characterized in terms of the entropy per unit time of the Poisson suspension.  The time evolution of the system can be monitored by observing the position and momentum of every particle incident on both sides of the wall at $z=0$ on a large but finite area $\Sigma$ including the small hole $\sigma$.  The surface $\Sigma$ is partitioned in small cells $\Delta^2A=\Delta x\Delta y$, which play the role of detectors measuring the position and momentum of every particle with a given resolution $\Delta^3 r \, \Delta^3p$.  The knowledge of all these events allows us to reconstruct the trajectories of all the particles in the gas since the particles move in free flight possibly interrupted by elastic collisions on the wall, as shown in Fig.~\ref{fig1}.  The number of particles incident on the cell $[{\bf r}_i, {\bf r}_i+\Delta{\bf r}]$ with the momentum $[{\bf p}_i, {\bf p}_i+\Delta{\bf p}]$ during the time interval $[0,T]$ is equal to
\be
N_i = \left\vert\frac{p_{zi}}{m}\right\vert \, T \; \Delta^2A \; F({\bf r}_i,{\bf p}_i) \; \Delta^3p \, ,
\ee
where $F({\bf r}_i,{\bf p}_i)$ is the distribution function (\ref{distrib_fn}) for the orbit $C$ followed by the corresponding particles.  The initial conditions of these $N_i$ particles are distributed uniformly in space inside the volume $\left\vert p_{zi}/m\right\vert T \Delta^2A$.  If space is discretized in cells of volume $\Delta x\Delta y\Delta z$, the number of different positions is given by
\be
M_i = \left\vert\frac{p_{zi}}{m}\right\vert \frac{T \; \Delta^2A}{\Delta x\Delta y\Delta z} =\left\vert\frac{p_{zi}}{m}\right\vert \frac{T}{\Delta z} \, .
\ee
The entropy per unit time is calculated as the rate of exponential growth with the time interval $T$ of the number of possible configurations of the $N_i$ particles in the $M_i$ positions \cite{G91,G92,G94}
\be
h = \lim_{T\to\infty} \frac{1}{T}\, \ln \prod_i \frac{M_i^{N_i}}{N_i!} \, .
\ee
Since the particles incident on each side of the surface $\Sigma$ are distributed at the temperature and density of the corresponding reservoir, the entropy per unit time is obtained as
\be
h = \int_{p_z>0} d^3p \int_{\Sigma} d^2A \left\vert\frac{p_{z}}{m}\right\vert F_{\rm L}({\bf p}) \, \ln \frac{\rm e}{F_{\rm L}({\bf p})\, \Delta^3 r \, \Delta^3p} + \int_{p_z<0} d^3p \int_{\Sigma} d^2A\left\vert\frac{p_{z}}{m}\right\vert F_{\rm R}({\bf p}) \, \ln \frac{\rm e}{F_{\rm R}({\bf p})\,\Delta^3 r \, \Delta^3p} \, ,
\label{dyn-entr-effusion}
\ee
where $F_{\rm L}({\bf p})$ and $F_{\rm R}({\bf p})$ are the single-particle distribution functions (\ref{distrib_fn}) respectively of the left-hand and right-hand reservoirs.  This entropy per unit time gives the accumulation rate of information by the detectors monitoring the process. Since these detectors have a finite resolution $\Delta^3 r \, \Delta^3p$, the accumulation rate of information depends on this resolution.  The fact that the entropy per unit time increases logarithmically with the resolution means that the random process is continuous in the corresponding variables $({\bf r},{\bf p})$ since the initial conditions of the particles are distributed uniformly in space at given density and according to a Maxwell-Boltzmann distribution in momentum.

Now, the coentropy per unit time can be calculated similarly taking into account that the orbits fall in the two subsets (\ref{C=L}) and (\ref{C=R}).  The time-reversal symmetry exchanges the values of the temperature and density only for the orbits going through the small hole of area $\sigma$ in the wall $W$.  For these orbits, the single-particle distribution function appearing in the logarithm should have the temperature and density of the time-reversed orbit.  However, the distribution function outside the logarithm should remain the same because the statistics continues to be carried out over the typical orbits of the process (and not their time reversals).  With these considerations, the time-reversed coentropy per unit time is given by
\bea
h^{\rm R} &=& \int_{p_z>0} d^3p \left[\int_{\Sigma\setminus\sigma} d^2A \left\vert\frac{p_{z}}{m}\right\vert F_{\rm L}({\bf p}) \, \ln \frac{\rm e}{F_{\rm L}({\bf p})\, \Delta^3 r \, \Delta^3p} +\int_{\sigma} d^2A \left\vert\frac{p_{z}}{m}\right\vert F_{\rm L}({\bf p}) \, \ln \frac{\rm e}{F_{\rm R}({\bf p})\, \Delta^3 r \, \Delta^3p} \right] \nonumber\\
&&+ \int_{p_z<0} d^3p \left[\int_{\Sigma\setminus\sigma} d^2A \left\vert\frac{p_{z}}{m}\right\vert F_{\rm R}({\bf p}) \, \ln \frac{\rm e}{F_{\rm R}({\bf p})\,\Delta^3 r \, \Delta^3p}+\int_{\sigma} d^2A \left\vert\frac{p_{z}}{m}\right\vert F_{\rm R}({\bf p}) \, \ln \frac{\rm e}{F_{\rm L}({\bf p})\,\Delta^3 r \, \Delta^3p}\right] .
\label{dyn-coentr-effusion}
\eea

Taking the difference between the coentropy (\ref{dyn-coentr-effusion}) and the entropy (\ref{dyn-entr-effusion}), we find the contributions of the orbits for which the Poisson measure is not time-reversal symmetric:
\be
h^{\rm R}-h = \frac{1}{2} \int_{{\mathbb R}^3} d^3p \int_{\sigma} d^2A \left\vert\frac{p_{z}}{m}\right\vert \left[ F_{\rm L}({\bf p})-F_{\rm R}({\bf p})\right] \, \ln \frac{F_{\rm L}({\bf p})}{F_{\rm R}({\bf p})} \, ,
\label{coentr-entr-effusion}
\ee
where the symmetry $p_z\to -p_z$ of the Maxwell-Boltzmann distributions has been used.  The ratio of the single-particle distribution functions of both reservoirs can be written in terms of the affinities (\ref{AE})-(\ref{AN}) as
\be
\frac{F_{\rm L}({\bf p})}{F_{\rm R}({\bf p})} = \exp\left( \epsilon A_E + A_N\right) ,
\ee
where $\epsilon={\bf p}^2/(2m)$ is the kinetic energy of the particle.
On the other hand, the single-particle distribution functions multiplied by the velocity $\vert p_z/m\vert$ and integrated over some surface represent the fluxes of particles of given momentum through this surface.  Integrating these fluxes over the corresponding values of the momentum gives the transition rates (\ref{rate-L}) and (\ref{rate-R}).  Accordingly, we get
\be
h^{\rm R}-h =\int_0^{\infty} d\epsilon \left[ w_{\rm L}(\epsilon)-w_{\rm R}(\epsilon)\right] \left( \epsilon A_E + A_N\right) = A_E \, \langle J_E\rangle + A_N \, \langle J_N\rangle = \frac{1}{k_{\rm B}} \frac{d_{\rm i}S}{dt}
\ee
in terms of the average values (\ref{JE}) and (\ref{JN}) of the energy and particle currents and, thus, the thermodynamic entropy production in the nonequilibrium steady state.  Therefore, the breaking of the time-reversal symmetry by the nonequilibrium steady state is characterized by the thermodynamic entropy production.

\subsection{Quantum systems}

We may wonder if such considerations would extend to quantum systems.  Let us take electrons in a one-dimensional lattice \cite{BB00,T01,JP01,JP02,AJPP07,BJP13}.  The Hamiltonian operator ruling this system reads
\be
H = -\sum_{l=-\infty}^{+\infty} \gamma_l \left( d_l^{\dagger} \, d_{l+1} + d_{l+1}^{\dagger} \, d_{l} \right) ,
\ee
where $d_l^{\dagger}$ and $d_l$ are anticommuting creation-annihilation operators on lattice sites $l\in{\mathbb Z}$ separated by the distance~$a$.  We consider a single spin orientation in order to simplify the notations and the discussion. The tunneling amplitudes are supposed to become asymptotically constant
\be
\lim_{l\to\pm\infty} \gamma_l = \gamma
\ee
in order to model the scattering of electrons on a perturbation localed at finite distance.  The resolution of this scattering problem is well known in terms of the coefficients of transmission $0\leq T(\epsilon)\leq 1$ and reflection $R(\epsilon)=1-T(\epsilon)$ for incident waves coming from the left-hand or right-hand sides of the scatterer with the energy $\epsilon=-2\gamma\cos(pa/\hbar)$ where $p$ is the momentum of the electron and $\hbar$ Planck's constant \cite{BB00,T01,JP01,JP02,AJPP07,BJP13}.  The left-hand and right-hand semi-infinite sides of the scatterer constitute reservoirs.  The situation is similar to the classical effusion process.  However, the electrons are here distributed according to Fermi-Dirac distributions
\be
f_j=f_j(\epsilon) = \frac{1}{{\rm e}^{\beta_j(\epsilon-\mu_j)}+1} \qquad\mbox{with}\quad j={\rm L},{\rm R} \, ,
\label{F-D}
\ee
where $\epsilon=\epsilon(p)$ is the energy of an electron in the reservoirs, $\beta_j=(k_{\rm B}T_j)^{-1}$ is the inverse temperature of the $j^{\rm th}$~reservoir, and $\mu_j$ its chemical potential.

An entropy per unit time has been introduced for quantum systems by Connes, Narnhofer, and Thirring \cite{CNT87}.  In the case of quasi-free algebra, an explicit expression is known for this dynamical entropy \cite{NT87,BHK98}.  If the present system is considered in a nonequilibrium steady state, the electrons are freely coming from both reservoirs with their corresponding momentum~$p$ so that the entropy per unit time is here given by
\be
h = \int_0^{+\infty} \frac{dp}{2\pi\hbar} \left\vert\frac{d\epsilon}{dp}\right\vert \left[ -f_{\rm L} \, \ln f_{\rm L} - (1-f_{\rm L}) \, \ln(1-f_{\rm L})\right] + \int_{-\infty}^0 \frac{dp}{2\pi\hbar} \left\vert\frac{d\epsilon}{dp}\right\vert \left[ -f_{\rm R} \, \ln f_{\rm R} - (1-f_{\rm R}) \, \ln(1-f_{\rm R})\right] .
\ee
This expression is analogous to Eq.~(\ref{dyn-entr-effusion}) since $\vert d\epsilon/dp\vert$ is the electron velocity corresponding to the classical velocity $\vert p_z/m\vert$.  The logarithmic terms have the form consistent with Pauli exclusion principle according to which every electronic orbital is either empty or occupied by at most one electron.

A time-reversed coentropy per unit time can be defined by considering that the electrons reflected by the scatterer go back to the same reservoir at the same temperature and chemical potential, while the electrons transmitted through the scatterer go to the other reservoir so that, under time reversal, the Fermi-Dirac distributions in the logarithms remain the same with probability $R(\epsilon)$, but are replaced by the one of the opposite reservoir with probability $T(\epsilon)$.  Accordingly, the time-reversal coentropy per unit time can be written as
\bea
h^{\rm R} &=& \int_0^{+\infty} \frac{dp}{2\pi\hbar} \left\vert\frac{d\epsilon}{dp}\right\vert \left\{R(\epsilon)\left[ -f_{\rm L} \, \ln f_{\rm L} - (1-f_{\rm L}) \, \ln(1-f_{\rm L})\right]+T(\epsilon)\left[ -f_{\rm L} \, \ln f_{\rm R} - (1-f_{\rm L}) \, \ln(1-f_{\rm R})\right]\right\} \nonumber\\
&&+ \int_{-\infty}^0 \frac{dp}{2\pi\hbar} \left\vert\frac{d\epsilon}{dp}\right\vert \left\{R(\epsilon)\left[ -f_{\rm R} \, \ln f_{\rm R} - (1-f_{\rm R}) \, \ln(1-f_{\rm R})\right]+T(\epsilon)\left[ -f_{\rm R} \, \ln f_{\rm L} - (1-f_{\rm R}) \, \ln(1-f_{\rm L})\right]\right\} .
\eea
Since the dispersion relation giving the energy has the symmetry $\epsilon(-p)=\epsilon(p)$ and the Fermi-Dirac distributions (\ref{F-D}) only depend on the energy $\epsilon$, the difference between the coentropy and the entropy is equal to
\be
h^{\rm R}-h = \int_0^{\infty} \frac{d\epsilon}{2\pi\hbar} \, T(\epsilon) \left( f_{\rm L}-f_{\rm R}\right) \ln\frac{f_{\rm L}(1-f_{\rm R})}{f_{\rm R}(1-f_{\rm L})} \, .
\ee
Because the Fermi-Dirac distributions (\ref{F-D}) satisfy
\be
\frac{f_j}{1-f_j} = {\rm e}^{-\beta_j(\epsilon-\mu_j)} \qquad \mbox{for}\qquad j={\rm L},{\rm R} \, ,
\ee
we find again that 
\be
h^{\rm R}-h = A_E \, \langle J_E\rangle + A_N \, \langle J_N\rangle = \frac{1}{k_{\rm B}} \frac{d_{\rm i}S}{dt}
\label{hR-h-Qthermo}
\ee
in terms of the affinities (\ref{AE}) and (\ref{AN}) but with the average currents here given by the Landauer-B\"uttiker formulas
\bea
&& \langle J_E\rangle = \int_0^{\infty} \frac{d\epsilon}{2\pi\hbar} \, \epsilon \, T(\epsilon) \left( f_{\rm L}-f_{\rm R}\right) \, , \label{JE-F}\\
&& \langle J_N\rangle  = \int_0^{\infty} \frac{d\epsilon}{2\pi\hbar} \, T(\epsilon) \left( f_{\rm L}-f_{\rm R}\right)\, , \label{JN-F}
\eea
for a single spin orientation \cite{BB00,T01,JP01,JP02,AJPP07,BJP13}.
Here also, the difference between the coentropy and the entropy per unit time is related to the thermodynamic entropy production.

The correspondence between the quantum and classical formulas is obtained by fixing the arbitrary phase-space volume introduced in the classical framework thanks to the non-vanishing quantum value of Planck's constant according to $\Delta^3r\Delta^3p=(2\pi\hbar)^3$ \cite{G91,G92,G94}.

Relationships similar to Eq.~(\ref{hR-h-Qthermo}) involving a relative entropy at the path level of description and characterizing time-reversal symmetry breaking have been obtained in a related framework \cite{CDJMN04}.

\section{Statistics of current fluctuations}
\label{Currents}

The breaking of the time-reversal symmetry by the nonequilibrium steady state also manifests itself at the level of the fluctuations of the currents flowing across the system.  This is nicely expressed in the so-called current fluctuation theorems which have been proved for stochastic processes as well as for open quantum systems \cite{AG04,AG06JSM,AG07JSM,AG07JSP,FB08,SU08,AGMT09,EHM09,TH07,TLH07,CHT11}.

\subsection{Current fluctuation theorem for stochastic processes}

A Markovian process ruled by the master equation (\ref{master}) can be schematically represented by a graph \cite{S76}.  The vertices of the graph correspond to the coarse-grained states $\{\omega\}$ and the edges to the possible transitions $\omega\leftrightharpoons \omega'$.  The graph can be decomposed into cycles, allowing the definition of the instantaneous currents ${\bf j}(t)$ and the identification of the affinities $\bf A$ driving the system out of equilibrium \cite{AG07JSP,S76}.  These affinities are for instance the thermal affinity~(\ref{AE}), as well as the chemical affinities~(\ref{AN}) for the different particle species that may flow across the system.  Their fluctuating currents are defined by averaging the instantaneous currents over some given time interval $[0,t]$ as
\be
{\bf J} = \frac{1}{t} \int_0^t {\bf j}(t') \, dt' \, .
\ee
The system is supposed to evolve in a nonequilibrium steady state of affinities $\bf A$.  If $P_{\bf A}({\bf J})$ denotes the probability density that the fluctuating currents would take the values $\bf J$ in the steady state $\bf A$, the {\it current fluctuation theorem} asserts that the ratio of the probabilities of opposite fluctuations behaves as
\be
\frac{P_{\bf A}({\bf J})}{P_{\bf A}(-{\bf J})}\simeq {\rm e}^{{\bf A}\cdot{\bf J}\, t} \qquad\mbox{for}\qquad t\to \infty \, .
\label{FT}
\ee
At equilibrium where the affinities vanish, the principle of detailed balance is recovered because the probabilities of the opposite fluctuations are thus in balance: $P_{\bf 0}({\bf J})\simeq P_{\bf 0}(-{\bf J})$.  Out of equilibrium, the fluctuation theorem expresses the directionality induced by the affinities on the current fluctuations.

The current fluctuation theorem can be alternatively expressed in terms of the cumulant generating function
\be
Q_{\bf A}(\pmb{\lambda}) = \lim_{t\to\infty} - \frac{1}{t} \ln \left\langle {\rm e}^{-\pmb{\lambda}\cdot{\bf J} \, t}\right\rangle_{\bf A}
\label{Q}
\ee
which is also a large-deviation property \cite{T09}.  The average values of the currents, their diffusivities, as well as their higher cumulants are given by taking the successive derivatives of this generating function with respect to the corresponding counting parameters $\pmb{\lambda}$:
\bea
\langle J_{\alpha}\rangle_{\bf A} &=& \frac{\partial Q_{\bf A}}{\partial
\lambda_{\alpha}}\Big\vert_{\pmb{\lambda}={\bf 0}} \, , \label{av_J} \\
D_{\alpha\beta}({\bf A}) &=& - \frac{1}{2} \frac{\partial^2 Q_{\bf A}}{\partial
\lambda_{\alpha}\partial \lambda_{\beta}}\Big\vert_{\pmb{\lambda}={\bf 0}}\, , \label{D}\\
C_{\alpha\beta\gamma}({\bf A}) &=& \frac{\partial^3 Q_{\bf A}}{\partial
\lambda_{\alpha}\partial \lambda_{\beta}\partial \lambda_{\gamma}}\Big\vert_{\pmb{\lambda}={\bf 0}} \, , \label{C}\\
&\vdots& \nonumber
\eea

The remarkable result is that the generating function (\ref{Q}) obeys the symmetry relation
\be
Q_{\bf A}(\pmb{\lambda}) = Q_{\bf A}({\bf A}-\pmb{\lambda}) \, ,
\label{FT-Q}
\ee
as the direct consequence of the fluctuation theorem (\ref{FT}).  This symmetry has implications when the generating function is differentiated with respect to the counting parameters $\pmb{\lambda}$ and the affinities $\bf A$.  For this reason, the fluctuation theorem has consequences among the statistical cumulants of the currents and their derivatives with respect to the affinities, which probe the response properties of the system.  Indeed, an average current can be expanded in powers of the affinities as
\be
\langle J_\alpha\rangle_{\bf A} = \sum_\beta L_{\alpha,\beta} \, A_{\beta} + \frac{1}{2} \sum_{\beta,\gamma} M_{\alpha,\beta\gamma} \, A_{\beta}  \, A_{\gamma} + \cdots
\ee
with the linear and nonlinear response coefficients:
\bea
L_{\alpha,\beta} &=& 
\frac{\partial^2 Q_{\bf A}}{\partial \lambda_{\alpha}\partial A_{\beta}}\Big\vert_{\pmb{\lambda}={\bf A}={\bf 0}} \, ,\\
M_{\alpha,\beta\gamma} &\equiv& 
\frac{\partial^3 Q_{\bf A}}{\partial \lambda_{\alpha}\partial A_{\beta}\partial A_{\gamma}}\Big\vert_{\pmb{\lambda}={\bf A}={\bf 0}} \, ,\label{M}\\
&& \nonumber \\
&\vdots& \nonumber
\eea
The symmetry relation (\ref{FT-Q}) not only implies the Green-Kubo formulas and the Onsager reciprocity relations for the linear response coefficients:
\bea
L_{\alpha,\beta} &=& D_{\alpha\beta}({\bf 0}) \, , \\
L_{\alpha,\beta} &=&  L_{\beta,\alpha} \, ,
\eea
but also generalizations of these relations to the higher cumulants and the nonlinear response coefficients \cite{AG07JSM,BK77,BK79}.  In particular, the nonlinear response coefficients (\ref{M}) are related to the diffusivities (\ref{D}) according to
\be
M_{\alpha,\beta\gamma} = \left(\frac{\partial D_{\alpha\beta}}{\partial A_{\gamma}}+
\frac{\partial D_{\alpha\gamma}}{\partial A_{\beta}}\right)_{{\bf A}={\bf 0}} \, ,
\ee
which is a generalization of the Green-Kubo formulas.  A generalization of the Onsager reciprocity relations is given by
\be
\left(\frac{\partial C_{\alpha\beta\gamma}}{\partial A_\delta}\right)_{{\bf A}={\bf 0}} = -\frac{1}{2}\,\frac{\partial^4 Q_{\bf A}}{\partial\lambda_{\alpha}\partial\lambda_{\beta}\partial\lambda_{\gamma}\partial \lambda_{\delta}}\Big\vert_{\pmb{\lambda}={\bf A}={\bf 0}} \, ,
\ee
which shows that the left-hand side is totally symmetric under the permutations of the four indices.
Similar relations exist at higher orders as well.  They are the consequences of the underlying microreversibility.

Nowadays, the current fluctuation theorem and its consequences for the linear and nonlinear response coefficients have been established for a broad range of stochastic processes, in particular, describing nonequilibrium chemical reactions \cite{AG04,AG08JCP}, molecular motors \cite{AG06PRE,LM09}, effusion processes \cite{CVK06,GA11}, diffusion processes \cite{HPPG11}, gas flows ruled by the fluctuating Boltzmann equation \cite{G13}, and mesoscopic electronic circuits with quantum dots \cite{AG06JSM,BEG11}.

\subsection{Current fluctuation theorem for open quantum systems}

Time-reversal symmetry relations have also been established for the currents flowing across open quantum systems driven out of equilibrium by non-vanishing affinities \cite{AG06JSM,FB08,SU08,AGMT09,EHM09,TH07,TLH07,CHT11,TN05}.  For quantum systems, these relations have been established directly from the Hamiltonian quantum dynamics of the system in contact with the reservoirs.  In the presence of an external magnetic field $B$, the symmetry of a time-dependent Hamiltonian operator $H(t;B)$ under time reversal $\Theta$ reads
\be
\Theta\, H(t; B)\, \Theta^{-1} = H(t; - B) \, .
\label{TR}
\ee
The external magnetic field is reversed because the external currents in the coils generating the field are also reversed under time reversal.

The statistics of the currents is carried out between two measurements separated by some time interval $[0,t]$ \cite{AGMT09}.  $\Delta E_j$ and $\Delta N_{jk}$ denote the variations in the energy and the number of particles $k$ in the $j^{\rm th}$ reservoir during the time interval $[0,t]$.  A forward protocol is compared with a reversed protocol.  The forward and reversed protocols start with the reservoirs in grand-canonical equilibrium states at the inverse temperatures $\beta_j$ and chemical potentials $\mu_{jk}$.  Comparing the probabilities of opposite variations of the energies and particles numbers during the forward and reversed protocols, the following relation is the consequence of the time-reversal symmetry (\ref{TR})
\be
\frac{P_{\rm F} (\Delta E_j,\Delta N_{jk};B)}
{P_{\rm R} (-\Delta E_j,-\Delta N_{jk};-B)}
 ={\rm e}^{\sum_j\beta_j(\Delta E_{j}-\sum_{k}\mu_{jk}\Delta N_{jk}-\Delta\Phi_{j})} 
\label{PFT}
\ee
in terms of the differences $\Delta\Phi_{j}$ in the grand potentials during the time interval $[0,t]$ \cite{AGMT09,EHM09,TH07,TLH07,CHT11}.  In the case of a closed system driven by a time-dependent external force, this relation is known as quantum Crooks fluctuation theorem and it implies quantum Jarzynski nonequilibrium work equality \cite{TH07,TLH07,CHT11}.  This latter can also be obtained from a different approach based on quantum functional symmetry relations \cite{AG08PRL}.

In any case, the relation (\ref{PFT}) concerns a transient process driven by the time-dependent Hamiltonian $H(t;B)$.  If we are interested in the current fluctuation theorem for the system in nonequilibrium steady states, the Hamiltonian should remain essentially constant during the time interval $[0,t]$.  Moreover, the symmetry relation (\ref{PFT}) should be transformed into a new relation for the {\it differences} of energies and particle numbers flowing between the $j^{\rm th}$~reservoir and some reference reservoir.  Indeed, the nonequilibrium process is driven by the affinities (\ref{AE})-(\ref{AN}) that are defined by the {\it differences} of inverse temperatures and chemical potentials with respect to the reference reservoir.  Nevertheless, a current fluctuation theorem can be proved for open quantum systems in nonequilibrium steady states of given affinities~$\bf A$ and the symmetry relation
\be
Q_{\bf A}(\pmb{\lambda};B) = Q_{\bf A}({\bf A}-\pmb{\lambda};-B)
\label{FT-Q-B}
\ee
has been obtained for the cumulant generating function of the currents in the presence of an external magnetic field \cite{AGMT09}.  The theory leads to the Levitov-Lesovik formula for the generating function of full counting statistics in electron quantum transport \cite{LL93,LLL96,G13SmallBook}.

By taking successive derivatives with respect to the counting parameters $\pmb{\lambda}$, the Green-Kubo formulas and Casimir-Onsager reciprocity relations can be generalized to higher-order responses properties thanks to the current fluctuation theorem \cite{SU08,AGMT09,BK77,BK79}.  These results are investigated experimentally in  mesoscopic electronic circuits \cite{UGMSFS10,NYHCKOLESUG10,NYHCKOLESUG11,KRBMGUIE12}. 

\section{Equilibrium systems}
\label{Equil}

Broken symmetries in equilibrium states can be characterized by using similar considerations as for the breaking of time reversal in nonequilibrium steady states \cite{G12JSM,G12PS}.  In magnetic systems at equilibrium for instance, a non-vanishing magnetization can be induced by an external magnetic field, which thus breaks the symmetry of the Hamiltonian.

Let us consider systems composed of $N$ spins $\pmb{\sigma}=\{\sigma_i\}_{i=1}^N$ with $\sigma_i\in\{+1,-1\}$.  The energy of the system is given by the Hamiltonian function 
\be
H_N(\pmb{\sigma};B) = H_N(\pmb{\sigma};0)- B \, M_N(\pmb{\sigma}) \, ,
\label{H}
\ee
where $B$ is the external magnetic field and
\be
M_N(\pmb{\sigma})=\sum_{i=1}^N \sigma_i
\label{Magn}
\ee
the magnetization, which is the order parameter.  The {\it spin reversal} is defined as
\be
\pmb{\sigma}^{\rm R}=R\, \pmb{\sigma}=-\pmb{\sigma} \, ,
\label{spin-rev}
\ee
which generates the discrete group ${\mathbb Z}_2=\{ 1,R\}$.  

In the absence of external magnetic field, the Hamiltonian is symmetric under spin reversal while the magnetization is reversed:
\bea
R \, H_N(\pmb{\sigma};0) \, R &=& H_N(\pmb{\sigma};0) \, ,  \label{H-sym}\\
R \, M_N(\pmb{\sigma}) \, R &=& - M_N(\pmb{\sigma}) \, ,\label{M-sym}
\eea
so that
\be
R \, H_N(\pmb{\sigma};B) \, R = H_N(\pmb{\sigma};-B) \, .
\label{H-B}
\ee

The system is supposed to be in the Gibbsian canonical equilibrium state (\ref{Gibbs}) at the temperature $T$ with $E_{\pmb{\sigma}}=H_N(\pmb{\sigma};B)$.  Examples of such systems are given by the Ising and Curie-Weiss models. In these systems, the external field is known to induce a magnetization, which breaks the ${\mathbb Z}_2$ symmetry of the Hamiltonian in the absence of external field.

\subsection{Entropy, coentropy, and broken symmetry}

For such equilibrium systems, we can define the usual thermodynamic entropy
\be
S = -k_{\rm B} \sum_{\pmb{\sigma}} p_{\pmb{\sigma}} \ln p_{\pmb{\sigma}} \, ,
\ee
as well as the spin-reversed coentropy
\be
S^{\rm R} = -k_{\rm B} \sum_{\pmb{\sigma}} p_{\pmb{\sigma}} \ln p_{\pmb{\sigma}^{\rm R}}
\ee
in analogy with Eq.~(\ref{dyn-coentr}).  Their difference is proportional to the Kullback-Leibler divergence
\be
D\left(p_{\pmb{\sigma}}\Vert p_{\pmb{\sigma}^{\rm R}}\right) = \sum_{\pmb{\sigma}} p_{\pmb{\sigma}} \ln \frac{p_{\pmb{\sigma}}}{p_{\pmb{\sigma}^{\rm R}}} = \frac{1}{k_{\rm B}} \, \left( S_N^{\rm R}-S_N\right) = 2 \, \frac{B \langle M_N\rangle_B}{k_{\rm B}T} \geq 0 \, .
\label{KL-div-eq}
\ee
Therefore, the average value of the magnetization is always oriented in the same direction as the external magnetic field under the assumptions (\ref{H-sym})-(\ref{M-sym}).  Moreover, the coentropy is always larger or equal to the entropy.   The relation (\ref{KL-div-eq}) is to spin reversal what Eqs.~(\ref{coentr-entr})-(\ref{KL-div}) are to time reversal \cite{G12JSM}.

\subsection{Fluctuation theorem for the magnetization}

In a large but finite system composed of $N$ spins, the magnetization is a fluctuating variable and we may introduce the probability $P_B(M)$ that the magnetization takes the value $M=M_N(\pmb{\sigma})$ as
\be
P_B(M) = \left\langle \delta_{M,M_N(\pmb{\sigma})}\right\rangle_B \, .
\ee
In systems with the symmetry (\ref{H-B}), this probability distribution obeys the fluctuation theorem:
\be
\frac{P_B(M)}{P_B(-M)} = {\rm e}^{2\beta B M}
\label{FT-spin}
\ee
with $\beta=(k_{\rm B}T)^{-1}$ \cite{G12JSM,G12PS}.  This relation is proved as follows:
\bea
P_B(M) &=& \frac{1}{Z_N(B)} \sum_{\pmb{\sigma}} {\rm e}^{-\beta H_N(\pmb{\sigma};0)+\beta B M_N(\pmb{\sigma})} \; \delta_{M,M_N(\pmb{\sigma})} \nonumber\\
&=&
\frac{1}{Z_N(B)} \sum_{\pmb{\sigma}} {\rm e}^{-\beta H_N(\pmb{\sigma};0)-\beta BM_N(\pmb{\sigma})} \; \delta_{M,-M_N(\pmb{\sigma})} \nonumber\\
&=&
\frac{1}{Z_N(B)}\; {\rm e}^{2\beta B M} \sum_{\pmb{\sigma}} {\rm e}^{-\beta H_N(\pmb{\sigma};0)+\beta BM_N(\pmb{\sigma})} \; \delta_{-M,M_N(\pmb{\sigma})} \nonumber\\
&=& {\rm e}^{2\beta B M} \; P_B(-M) \, ,
\eea
where the sum over all the spin configurations has been replaced by the equivalent sum over the spin-reversed configurations since they both form to the same set: $R\{+1,-1\}^{N}=\{+1,-1\}^{N}$.  Thereafter, the symmetries (\ref{H-sym}) and (\ref{M-sym}) are used.  Finally, the Kronecker delta function allows us to restore the canonical equilibrium distribution by factorizing $\exp(2\beta B M)$ out of the sum. 

Here, the fluctuation relation characterizes the breaking of the spin-reversal symmetry by the external magnetic field.  If this latter vanishes, we recover the symmetry $P_0(M)=P_0(-M)$.  Otherwise, the fluctuations of magnetization prefer the direction given by the external magnetic field.  The equilibrium fluctuation relation (\ref{FT-spin}) is analogous to the nonequilibrium fluctuation relation (\ref{FT}).

\section{Conclusions and perspectives}
\label{Conclusions}

In this paper, an overview is presented of recent advances in nonequilibrium statistical mechanics about the statistics of random paths and current fluctuations.  These advances are developed in an approach where the statistics of random events is performed in time or spacetime in analogy with equilibrium statistical mechanics where statistics is performed in space.  In this approach, large-deviation relationships can be established between nonequilibrium properties such as the transport coefficients, the thermodynamic entropy production, or the affinities, and quantities characterizing the underlying microscopic Hamiltonian dynamics or corresponding stochastic processes.  Several classes of systems have been studied in this way.

Open classical Hamiltonian systems can be considered in the escape-rate formalism \cite{GN90,G98}.  These systems are set in transient nonequilibrium regimes by allowing their trajectories to escape from some phase-space domain delimited by absorbing boundary conditions, which define first-passage problems.  A conditionally invariant probability measure can be constructed, which has for support the set of trapped trajectories that do not escape.  This probability measure determines the escape rate, the Kolmogorov-Sinai entropy per unit time -- which is non-vanishing in the case of chaotic dynamics -- and the Lyapunov exponents characterizing the sensitivity to initial conditions.  In systems sustaining transport by diffusion, viscosity, or heat conductivity, the escape rate is proportional to the corresponding transport coefficient and a geometrical factor.  In this formalism, early large-deviation relationships could be established between nonequilibrium properties and the characteristic quantities of chaos in the underlying dynamics \cite{GN90,G98}.  

These chaos-transport relationships have inspired further work on systems maintained in nonequilibrium steady states by their interaction with reservoirs.  The presence of many degrees of freedom in the reservoirs can be taken into account by describing such systems in terms of stochastic processes.  The system is driven out of equilibrium by thermodynamic forces, also called affinities, that are defined in terms of the difference of temperatures and chemical potentials between the reservoirs.  In nonequilibrium steady states, the invariant probability measure breaks the time-reversal symmetry and directionality appears in the form of average currents induced by the affinities and flowing across the system.  The thermodynamic entropy production turns out to characterize time-reversal symmetry breaking, as shown by large-deviation relationships similar to the chaos-transport formulas.  Indeed, the entropy production can be expressed as the difference between two quantities. The first, called coentropy, is the statistical average of the decay rate of the probabilities to find time-reversed paths in the process and the second is the analogue of the Kolmogorov-Sinai entropy for the stochastic process.  The difference gives a relative entropy, also called a Kullback-Leibler divergence, which is known to be always non negative so that the result is in agreement with the second law of thermodynamics.  The relationship is illustrated for effusion processes and extends to quantum systems with quasi-free fermionic particles.

Besides, large-deviation theory also applies to the statistics of current fluctuations in nonequilibrium steady states.  Multivariate fluctuation theorems have been proved for all the currents flowing across a system in contact with several reservoirs on the ground of microreversibility.  Current fluctuation theorems have been obtained for stochastic processes \cite{AG07JSP}, as well as open quantum systems \cite{AGMT09}.  These theorems allow us to generalize the Green-Kubo formulas and the Onsager or Casimir-Onsager reciprocity relations from linear to nonlinear response properties \cite{AG07JSM,AGMT09}.  These results apply in particular to electron transport in semiconducting mesoscopic devices, as well as to molecular motors, and experiments are under way to investigate these relations and their implications \cite{UGMSFS10,NYHCKOLESUG10,NYHCKOLESUG11,KRBMGUIE12,TWOKM11}.

Going back to equilibrium statistical mechanics, fluctuation theorems and other large-deviation relationships can be used to study other broken symmetries than time reversal \cite{G12JSM,G12PS}.  For instance, magnetic systems in an external magnetic field may have a non-vanishing average magnetization, which thus breaks spin-reversal symmetry in much the same way as the average currents flowing across a nonequilibrium system break time-reversal symmetry.  This approach opens new promising perspectives to understand broken symmetries and their consequences in equilibrium and nonequilibrium systems.

\begin{acknowledgments}
This research is financially supported by the Universit\'e Libre de Bruxelles and the Belgian Federal Government under the Interuniversity Attraction Pole project P7/18 ``DYGEST".
\end{acknowledgments}


\end{document}